\documentclass{Interspeech}

\interspeechcameraready

\title{Analyzing the Impact of Accent on English Speech: Acoustic and Articulatory Perspectives}

\author[affiliation={1}]{Gowtham}{Premananth}
\author[affiliation={1}]{Vinith}{Kugathasan}
\author[affiliation={1}]{Carol}{Espy-Wilson}

\affiliation{Institues for Systems Research, Department of Electrical \& Computer Engineering}{University of Maryland, College Park}{USA}

\email{gowtham8@umd.edu , vinith@umd.edu, espy@umd.edu}

\usepackage{comment}
\usepackage{multicol}
\usepackage{multirow}

\begin{document}

\maketitle

\begin{abstract}

Advancements in AI-driven speech-based applications have transformed diverse industries ranging from healthcare to customer service. However, the increasing prevalence of non-native accented speech in global interactions poses significant challenges for speech-processing systems, which are often trained on datasets dominated by native speech. This study investigates accented English speech through articulatory and acoustic analysis, identifying simpler coordination patterns and higher average pitch than native speech. Using eigenspectra and Vocal Tract Variable-based coordination features, we establish an efficient method for quantifying accent strength without relying on resource-intensive phonetic transcriptions. Our findings provide a new avenue for research on the impacts of accents on speech intelligibility and offer insights for developing inclusive, robust speech processing systems that accommodate diverse linguistic communities.
\end{abstract}

\section{Introduction}

In recent years, significant progress in areas like automatic speech recognition (ASR), natural language processing (NLP), and generative AI tools has led to a surge in the use of speech-based applications, like AI-powered Virtual Assistants \cite{aiassistant}, End-to-End ASR Systems \cite{asr}, Speech-to-Text Integration in Business \cite{tts}, Healthcare Applications \cite{health,sch_gow}, and Multimodal AI \cite{acosta2022multimodal}, which are transforming industries from healthcare to customer service. These applications have become integral to everyday activities, offering seamless and efficient communication tools. A key challenge has emerged for widely spoken languages like English with the increasing prevalence of non-native accented speech in global interactions \cite{pycha2024influence}. Non-native speakers contribute to a diverse linguistic environment, yet their unique speech patterns often remain underrepresented in developing speech-processing systems.

The underlying issue lies in the training data of these systems, which predominantly consist of native speech samples. This imbalance results in significant performance disparities when these systems encounter accented speech \cite{bias}. Specifically, non-native speakers frequently experience higher error rates and misunderstandings when using ASR and related tools. Recent studies have shown the prevalence of these biases in ASR systems trained on widely spoken languages \cite{i1,i2}.  Such biases not only undermine the user experience but also hinder the broader adoption of speech-based technologies in multilingual and multicultural contexts.

Addressing these challenges necessitates a comprehensive investigation into the differences between native and accented English speech. By examining these differences from both articulatory and acoustic perspectives, researchers can identify the key factors contributing to system biases. Such insights can pave the way for the development of more inclusive and equitable speech-processing methods, enhancing both accessibility and performance across diverse user populations. This study aims to bridge this gap by focusing on the articulatory and acoustic characteristics of accented English speech and their implications for speech technology design. 

Accented speech varies significantly depending on the specific accent, influencing English pronunciation in unique ways. As a result, a generalized analysis of how accents affect speech is insufficient. Instead, it is crucial to deeply understand the distinct impacts of individual accents. While prior studies \cite{l1,l2} have attempted to measure the strength of foreign accents in English, most of these analyses rely on phonetic transcriptions. This approach, though valuable, is both time-intensive and costly, particularly when applied to accented speech. To address these challenges, this study aims to quantify the strength of different accents using acoustic and articulatory features, which can be extracted easily, offering a practical alternative for studying accented speech and its impact.

In summary, this study makes several key contributions to the field of speech processing. First, it addresses the critical gap in understanding the articulatory and acoustic differences between native and accented English speech, which are often overlooked in current speech-processing systems. By utilizing easily extractable acoustic and articulatory features, the study proposes a cost-effective and efficient approach to quantify the impact of accents, bypassing the labor-intensive process of using phonetic transcription. These contributions aim to foster the development of  equitable and robust speech technologies that can serve linguistically diverse users, ultimately improving accessibility and user experience across global contexts.

\section{Datasets}
\noindent\textbf{CMU ARCTIC Dataset~\cite{d1}}: This corpus comprises recordings from four native English speakers (two male, two female) sampled at 16 kHz. Each speaker reads approximately 1,300 phonetically balanced English prompts (ARCTIC prompts). In this study, we use this dataset as the non-accented baseline.

\noindent\textbf{L2 ARCTIC Dataset~\cite{d2}}:
This dataset features recordings from 24 non-native English speakers, representing six native languages: Arabic, Hindi, Korean, Mandarin, Spanish, and Vietnamese with a sampling rate of 44.1 kHz. Each language includes two male and two female speakers who read sentences from the CMU ARCTIC prompts. The dataset includes manually annotated phonetic transcriptions and alignments, identifying three types of mispronunciation errors: substitutions, deletions, and additions.  This dataset serves as the accented counterpart to the CMU ARCTIC dataset.

\noindent\textbf{Speech Accent Archive~\cite{d3}}:
This dataset contains recordings from 2,140 speakers, covering both native and non-native English accents. The dataset spans 177 countries and 214 native languages, with each speaker reading the same passage. Speaker demographics, including gender (1,103 male, 1,037 female), are provided. Table \ref{tab:pitch_comparison} presents the gender distribution for both native and non-native speakers matching the L2 ARCTIC language groups. The dataset is sampled at 44.1 kHz.

\section{Data preprocessing \& Feature extraction}

As we obtained data from different sources they had to be preprocessed before feature extraction. Since the audio files from different datasets were sampled at different sampling rates, we downsampled all the audio to 16kHz. This choice reduces the data size, making computations more efficient.

In our work, Mel-Frequency Cepstral Coefficients (MFCCs) and pitch were used as the acoustic features. Librosa~\cite{acoustic1}, a Python package for music and audio analysis was used to derive these features.

\noindent\textbf{Mel-Frequency Cepstral Coefficients (MFCCs): }MFCCs are a set of coefficients that capture the shape of the power spectrum of an audio signal. They are useful for capturing timbral features and are commonly used in tasks like speech recognition, accent classification, accent conversion, and speech synthesis.

\noindent\textbf{Pitch: }Pitch refers to the perceived frequency of a sound and is an important feature in analyzing speech and music. Here, we used the probabilistic YIN (PYIN) \cite{acoustic2} algorithm to estimate the pitch. The PYIN algorithm is an improved version of the YIN algorithm where probabilistic modeling is used to identify non-pitched regions (noise/unvoiced segments). 

\noindent\textbf{Vocal Tract Variables (TVs): }Speech-based articulatory features TVs that estimate the constriction degree and location of different articulators in the vocal tract were used as articulatory features. An acoustic-to-articulatory Speech Inversion (SI) system\cite{c2} was used to extract the TVs. This SI system uses a deep learning model trained on real articulatory data \cite{westbury1994speech} paired with corresponding speech utterances. This system is designed to learn how to estimate actual articulatory data directly from speech input. During the training of the SI system speech is converted into pretrained self supervised feature representations (HuBERT \cite{hsu2021hubert}) and used as the input to predict TVs. Once trained, the model can generalize to any speech utterance—predicting the most likely articulatory movements,  the TV trajectories, that would have occurred during its production. The trained SI system estimates 6 TVs: Lip Aperture, Lip Protrusion, Tongue Tip Constriction Degree, Tongue Tip Constriction Location, Tongue Body Constriction Degree, and Tongue Body Constriction Location. Along with the 6 TVs, source TVs aperiodictity and periodicity were extracted using a Aperioidicity, Periodicity, Pitch detector \cite{appdetector}.

\section{Methodology}

After extracting the low-level acoustic and articulatory features, they were converted to high-level coordination features (correlation matrix) \cite{c3} structures using channel-delay correlation. For TV-based correlation matrix, the normalized delayed correlation between $TV_i$ and $TV_j$ delayed by $d$ frames $( r_{TV_i , TV_j}^d)$, is computed using Eq.\ref{eq:delay}.

\begin{equation}
\label{eq:delay}
r_{TV_i , TV_j}^d=\frac{\sum_{t=1}^{T-d}TV_i[t]TV_j[t+d]}{\sqrt{r_{TV_i , TV_i}^0 r_{TV_j , TV_j}^0}}
\end{equation}

T is the frame number,~$r_{TV_i , TV_i}^0$,~$r_{TV_j , TV_j}^0$ are autocorrelation for $TV_i$ and $TV_j$. Such correlations are building blocks in a channel-delay correlation matrix $(R_{TV_i , TV_j}^n)$ which repetitively samples correlations as follows:

\begin{equation}
(R_{TV_i , TV_j}^n)=\begin{bmatrix}
r_{TV_i , TV_j}^0 & .. & r_{TV_i , TV_j}^{n(p-1)} & ..&r_{TV_i , TV_j}^{n(P-1)}\\
.. & .. & .. & ..&..\\
r_{TV_i , TV_j}^{-n(p-1)} & .. & r_{TV_i , TV_j}^0 & ..&r_{TV_i , TV_j}^{n(p-1)}\\
.. & .. & .. & ..&..\\
r_{TV_i , TV_j}^{-n(P-1)} & .. & r_{TV_i , TV_j}^{-n(p-1)} & ..&r_{TV_i , TV_j}^0\\
\end{bmatrix} \in \mathbb{R}^{(P \times P)}
\end{equation}

where $n$ is the time scale and $p \in [1,2, .. , P]$ is the number of selected correlations. The $R_{TV_i , TV_j}^n$ then become the building blocks in the high-level TV-based correlation matrix $\mathfrak{R}_{TV}$ as follows in Eq.\ref{eq:corr}:

\begin{equation}
\label{eq:corr}
\mathfrak{R}_{TV}=\begin{pmatrix}
R^{n=1}\\
R^{n=3}\\
R^{n=5}\\
R^{n=7}
\end{pmatrix}
=\begin{pmatrix}
[R_{TV_i , TV_j}^{n=1}]_{i,j \in [1,2,..,K]}\\
[R_{TV_i , TV_j}^{n=3}]_{i,j \in [1,2,..,K]}\\
[R_{TV_i , TV_j}^{n=5}]_{i,j \in [1,2,..,K]}\\
[R_{TV_i , TV_j}^{n=7}]_{i,j \in [1,2,..,K]}
\end{pmatrix}
\in \mathbb{R}^{4KP \times KP}
\end{equation}

where $R^{n=1} = [R_{TV_i , TV_j}^{n=1}]_{i,j \in [1,2,..,K]}$ represents a $K \times K$ matrix with
each entry being $R_{TV_i , TV_j}^{n=1}$ , which is a $P \times P$ matrix at the time delay scale $n = 1$, as in Eq.(2).
In our case $\mathfrak{R}_{TV}$ adopts four different time delay scales, i.e. $n \in {1, 3, 5, 7} $, $P = 10$ and $K = 6$ as we focus on the six TVs. Likewise for the MFCC-based correlation matrix $\mathfrak{R}_{MFCC}$ the only change is $K=13$ as we focus on the 13 MFCCs that are extracted.

Eigenspectra were generated from correlation matrices through eigenvalue decomposition, with eigenvalues rank-ordered by magnitude. These eigenspectra were then used to analyze coordination patterns, following the approach of Seneviratne et al.\cite{c1}. Their study employed correlation matrices from speech samples to differentiate between healthy individuals and those with depression. To characterize gestural coordination patterns, they simulated three types of speech such as overly simplified, natural, and erratic. By extracting eigenspectra from these signals, they demonstrated how eigenvalue distributions reflect different temporal coordination styles, offering insights into speech dynamics. Their findings showed that in speech with simpler coordination, lower-ranked eigenvalues start high, drop into negative values, and then stabilize near zero, whereas erratic speech follows the opposite trend beginning with negative values, peaking positively, and then decreasing toward zero. Following this, we conducted a coordination analysis to examine how accented speech differs from non-accented speech.

Beyond analyzing the acoustic and articulatory differences between accented and native English speech, we also examined how various accents influence speech characteristics. This was achieved through a coordination analysis comparing different accented speakers to native English speakers. Additionally, we analyzed the average pitch across accents to understand how pitch variations correlate with different linguistic backgrounds.

Our initial experiments utilized the CMU ARCTIC and L2 ARCTIC datasets, chosen for their phonetically balanced utterances from both native and non-native speakers. Averaging results across all utterances mitigated biases from individual phonetic variations, ensuring a comprehensive representation of speech characteristics. However, due to the limited number of speakers in these datasets, we extended our coordination and pitch analyses using the Speech Accent Archive. This allowed us to validate whether findings from L2 ARCTIC generalized to a broader speaker population.

In this work, we also investigated on how the strength of foreign accents in English can be measured using these acoustic and articulatory features. In initial works on measuring the strength of foreign accents in English \cite{l3}, the differences between the pronunciations of native and non-native English speakers were quantified using Levenshtein distances of the mispronunciations found in the phonetic transcriptions of accented speech. As a subset of utterances in the L2 ARCTIC dataset was annotated with phonetic transcriptions, we used those transcriptions to calculate the Levenshtein distance for different accents and used it as a baseline to investigate whether the quantified acoustic and articulatory features can also be used to measure the strength of accents.

\section{Results and Discussion}
\vspace*{-3pt}
\begin{figure}[h!]
    \centering
    \includegraphics[width=80mm, height= 65mm]{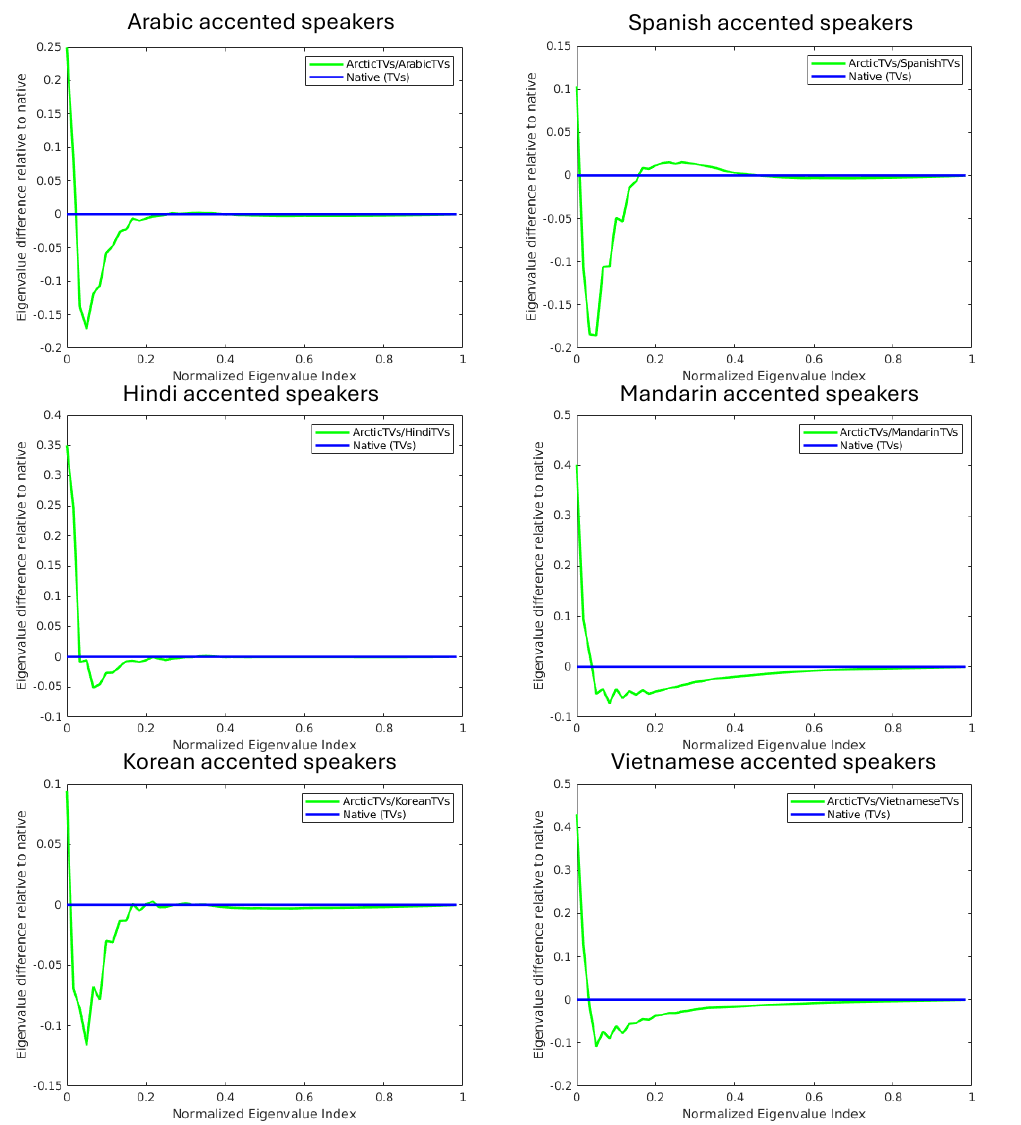}
    \caption{TV-based coordination analysis eigenspectra for accented speech when compared with native speech on the L2 ARCTIC dataset}
    \label{fig:tv}
\end{figure}
\begin{figure}[h!]
    \centering
    \includegraphics[width=80mm, height= 65mm]{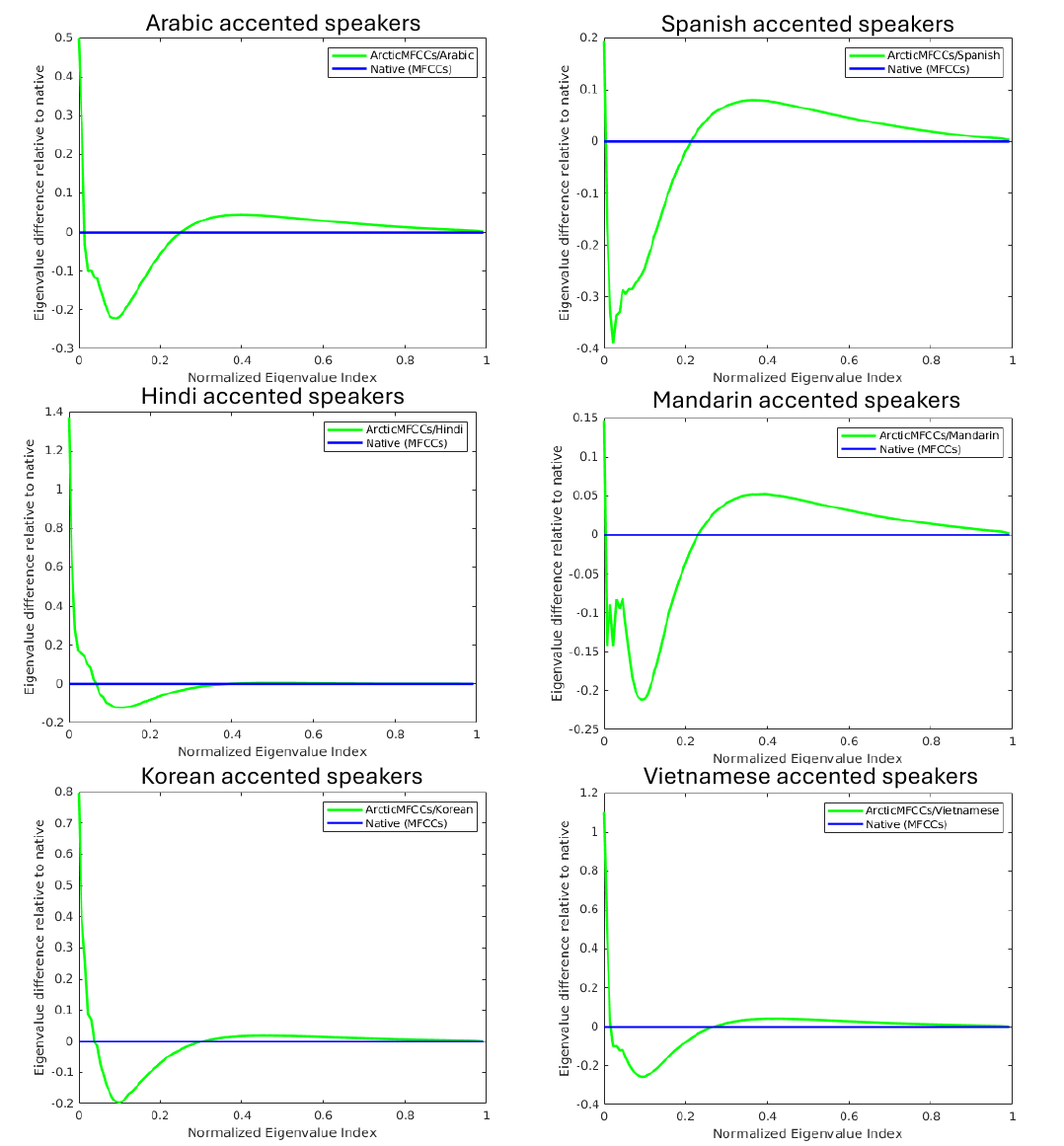}
    \caption{MFCC-based coordination analysis eigenspectra for accented speech when compared with native speech on the L2 ARCTIC dataset}
    \label{fig:mfcc}
\end{figure}

\vspace*{-3pt}
The TV-based coordination analysis done on different accented speakers from the L2 ARCTIC dataset when compared with native English speakers (Non-accented) from the CMU ARCTIC dataset produced eigenspectra that are shown in Fig.\ref{fig:tv}. The eigenspectra plots for the accented speech show that accented speech has simpler coordination when compared with non-accented native speech. This simpler coordination was seen across all the accents. The same trend of simpler coordination was observed from the eigenspectra derived from the acoustic features: MFCC-based correlation matrices obtained from accented speakers from the L2 ARCTIC dataset. These results are shown in Fig.\ref{fig:mfcc}.

\begin{figure}[h!]
    \centering
    \includegraphics[width=80mm, height= 65mm]{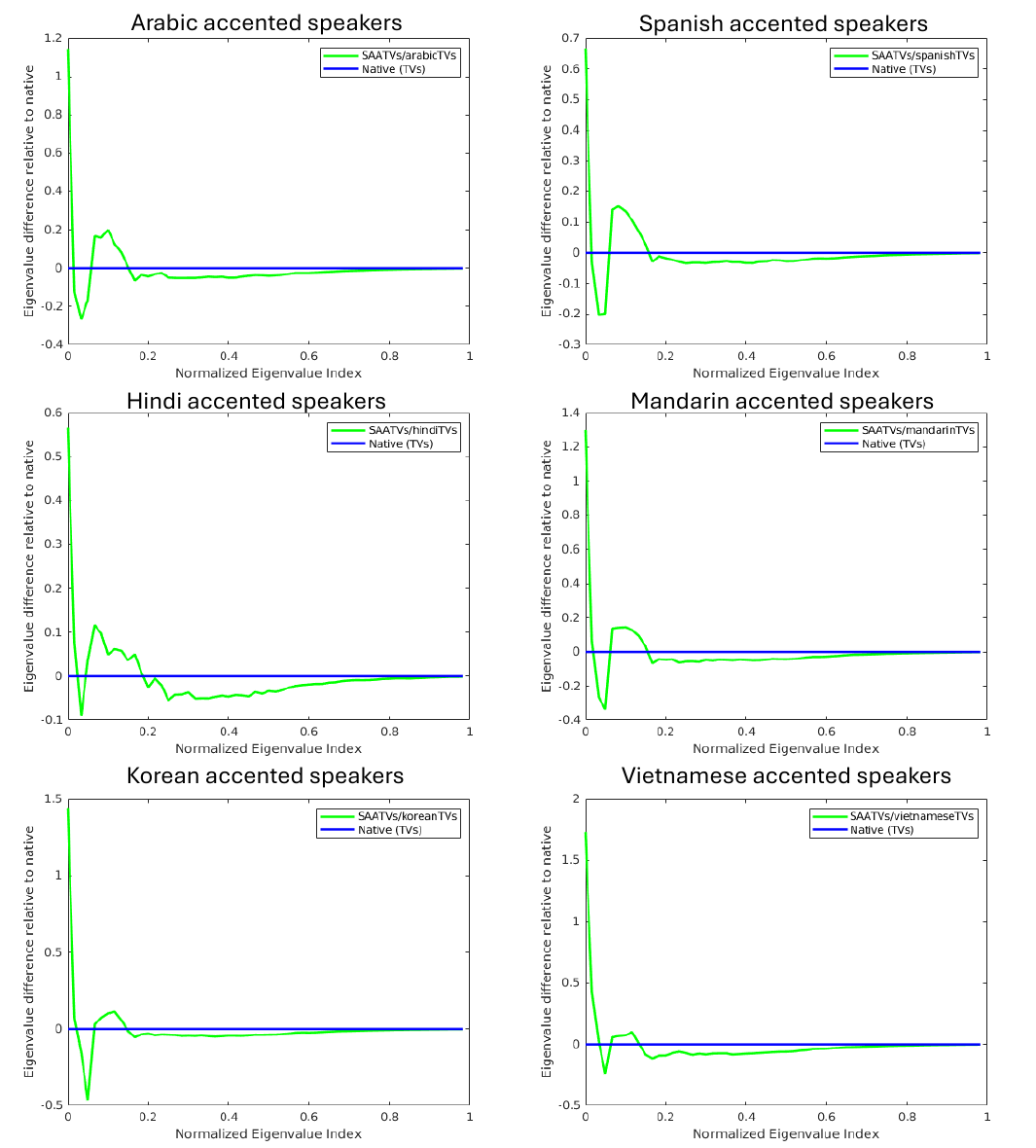}
    \caption{TV-based coordination analysis eigenspectra for accented speech when compared with native speech on the Speech accent archive dataset}
    \label{fig:sa_tv}
\end{figure}

The results of the TV-based coordination analysis done on the Speech accent archive dataset with accented speech from a large number of speakers reading a single paragraph are shown in Fig.\ref{fig:sa_tv}. This coordination analysis was also focused on the 6 languages found in the L2 ARCTIC dataset. The results show that even with a larger number of speakers the averaged eigenspectra shows simpler coordination for accented speech. 
\vspace*{-15pt}
\begin{table*}[th!]
\centering
\caption{Average pitch values across utterances and speakers for the selected first languages in the CMU ARCTIC, L2 ARCTIC, and Speech Accent Archive datasets. Throughout the table, the lowest average pitch values are marked in \textbf{bold}, and the second lowest values are \underline{underlined}.}

    \begin{tabular}{l c c c c c }
    \toprule
    \multirow{2}{*}{\textbf{Language}} & \multirow{2}{*}{\textbf{Gender}} & \multicolumn{2}{c}{\textbf{CMU\cite{d1} \& L2 ARCTIC\cite{d2}}} & \multicolumn{2}{c}{\textbf{Speech Archive\cite{d3}}} \\ 
    & & \textbf{\# Subjects} & \textbf{Pitch Avg (Hz)} & \textbf{\# Subjects} & \textbf{Pitch Avg (Hz)} \\ \midrule
    \multicolumn{6}{c}{\textbf{Non-accented Native speech}}\\ \midrule
    \multirow{2}{*}{English} & M & 2 & \textbf{110.58} & 309 & \underline{123.23} \\  
          & F & 2 & \textbf{179.36} & 270 & \textbf{196.79} \\ \midrule 
    \multicolumn{6}{c}{\textbf{Accented speech}}\\ \midrule     
    \multirow{2}{*}{Arabic} & M & 2 & 139.34 & 60 & 129.02 \\ 
           & F & 2 & 204 & 42 & 220.32 \\ \midrule 
    \multirow{2}{*}{Hindi} & M & 2 & 152.16 & 8 & 136.27 \\ 
             & F & 2 & 206.95 & 10 & 222.37 \\ \midrule 
    \multirow{2}{*}{Korean} & M & 2 & 128.24 & 24 & \textbf{121.76} \\  
          & F & 2 & 226.69 & 28 & \underline{200.43} \\ \midrule 
    \multirow{2}{*}{Mandarin} & M & 2 & 120.58 & 26 & 129.49 \\  
          & F & 2 & \underline{191.16} & 39 & 214.47 \\\midrule 
    \multirow{2}{*}{Spanish} & M & 2 & \underline{112.4} & 94 & 128.59 \\  
          & F & 2 & 195.24 & 68 & 205.51 \\ \midrule 
    \multirow{2}{*}{Vietanmese} & M & 2 & 160.17 & 13 & 125.76 \\  
          & F & 2 & 197.25 & 9 & 212.87 \\ \bottomrule
    
    \end{tabular}
\label{tab:pitch_comparison}
\end{table*}

While all accents exhibit simpler coordination patterns compared to native speech, notable differences emerge in their eigenspectra. The most significant variations are observed in the peak values of lower-ranked eigenvalues and the lowest points of negative-valued eigenvalues. These differences highlight how various accents influence speech with differing degrees of acoustic and articulatory changes.

Table \ref{tab:pitch_comparison} presents the results of the pitch analysis across multiple datasets, comparing native and accented speech. The findings indicate that speakers with accented speech consistently exhibit higher average pitch values than native English speakers. This trend is consistent across all datasets used in this study, reinforcing the impact of accents on pitch characteristics.
\begin{table}[h]
\centering
\caption{Strength of Accents quantified by Acoustic, Articulatory, and Phonetic measures in L2 ARCTIC dataset.}
\begin{tabular}{l c c c}
\toprule
\multirow{3}{*}{\textbf{Language}} & \multicolumn{3}{c}{\textbf{Strength of Accents}} \\ 
&\textbf{Articulatory} & \textbf{Acoustic} & \textbf{Levenshtein} \\ 
&\textbf{Coordination} & \textbf{Coordination} & \textbf{distance} \\  \midrule
Korean & 1.0227 & 1.8150 &8.1551 \\ 
Arabic & 1.4566 & 1.8921 &9.4124 \\ 
Hindi & 1.1578 & 2.6096 &10.7400 \\ 
Spanish & 1.3651 & 1.6293 &11.2383 \\
Mandarin & 1.3785 & 1.7237 &12.6433 \\ 
Vietnamese & 1.6295 & 1.7347 &16.4433 \\ \bottomrule
\end{tabular}
\label{tab:accent_comparison}
\end{table}

Quantized coordination scores for acoustic and articulatory analyses of accent strength were derived by computing the difference between coordination matrices of accented and native speech, followed by calculating the matrix norm of the averaged difference matrices for each accent. Table \ref{tab:accent_comparison} presents a comparative analysis of accent strength quantification methods, including acoustic, articulatory, and phonetic measures. Using Levenshtein distances from phonetic transcriptions as a baseline, TV-based articulatory coordination scores show a similar trend, with minor deviations in accent rankings, particularly for Arabic. Meanwhile, MFCC-based acoustic coordination scores maintain consistent rankings within subgroups (e.g., {Korean, Arabic, Hindi} and {Spanish, Mandarin, Vietnamese}) but deviate in overall accent ranking. These discrepancies likely arise because Levenshtein distances primarily capture mispronunciations, which represent only one aspect of accent strength. In contrast, acoustic and articulatory coordination features provide a more comprehensive measure by accounting for a broader range of accent-related influences, including their impact on speech intelligibility.

The analysis was limited by the availability of phonetic transcriptions for only a subset of utterances, restricting the study’s scope. Additionally, reliance on the L2 ARCTIC dataset, due to the absence of phonetic transcriptions in other datasets, limited the generalizability of the findings. Despite these constraints, results suggest that TV-based coordination features are a promising alternative to Levenshtein distances for quantifying accent strength. Future research should expand this work by incorporating a larger corpus with phonetic transcriptions and listener-rated intelligibility scores to better evaluate the effectiveness of TV-based coordination features.

\section{Conclusion and Future work}

The findings of this study highlight key distinctions in the articulatory and acoustic characteristics of accented versus native English speech. The eigenspectra derived from both TV-based and MFCC-based correlation matrices consistently reveal simpler coordination patterns in accented speech compared to non-accented native speech, across various datasets and accents. While these simpler patterns were observed universally, differences among accents were evident in the eigenspectra, particularly in the behavior of lower-ranked eigenvalues.

Pitch analysis further corroborated these trends, showing higher average pitch values for accented speakers across all datasets. Additionally, a comparative analysis of methods for quantifying accent strength demonstrated the potential of TV-based coordination features as a robust alternative to phonetic transcription-based Levenshtein distances. Unlike Levenshtein distances, which focus narrowly on mispronunciations, articulatory and acoustic features offer a more comprehensive view, capturing broader effects on intelligibility \& coordination.

Despite constraints like limited datasets and phonetic transcriptions, these results underscore the promise of articulatory coordination features in accent analysis. Future work should expand the dataset scope, include subjective intelligibility scores, and explore additional accents to refine and validate these findings. By advancing methods to quantify accent strength, this research contributes to the development of more inclusive and effective speech-processing technologies.
\bibliographystyle{IEEEtran}
\bibliography{template}

\end{document}